\documentclass[10pt]{iopart}  

\usepackage{cite}
\usepackage[english]{babel}
\usepackage[utf8]{inputenc} 
\usepackage[table]{xcolor} 
\usepackage[%
  colorlinks=true,
  urlcolor=blue,
  linkcolor=blue,
  citecolor=blue
]{hyperref}
\usepackage{algorithm} 
\usepackage{algpseudocode} 
\usepackage{bbold} 
\usepackage{amssymb}
\usepackage{bm}
\usepackage{bbm}
\usepackage{tabularx, booktabs}
\usepackage{xspace}
\usepackage{listings}
\usepackage{lipsum}
\usepackage{chemformula} 
\usepackage{physics}  

\newcolumntype{Y}{>{\centering\arraybackslash}X}
\definecolor{dgreen}{rgb}{0,.5,0}
\definecolor{dblue}{rgb}{0,0,.5}
\definecolor{dred}{rgb}{0.5,0,.5}

\usepackage{soul}
 
\begin{document}

\title{A Two-Step Rayleigh-Schrödinger Brillouin-Wigner Approach to Transition Energies}

 \author{Loris Delafosse} 
\address{Laboratoire de Chimie Quantique, Institut de Chimie,
CNRS/Université de Strasbourg, 4 rue Blaise Pascal, 67000 Strasbourg, France}
 
 \author{Amr Hussein} 
\address{Laboratoire de Chimie Quantique, Institut de Chimie,
CNRS/Université de Strasbourg, 4 rue Blaise Pascal, 67000 Strasbourg, France}

\author{Saad Yalouz}
\address{Laboratoire de Chimie Quantique, Institut de Chimie,
CNRS/Université de Strasbourg, 4 rue Blaise Pascal, 67000 Strasbourg, France}

\author{Vincent Robert} 
\ead{vrobert@unistra.fr}
\address{Laboratoire de Chimie Quantique, Institut de Chimie,
CNRS/Université de Strasbourg, 4 rue Blaise Pascal, 67000 Strasbourg, France}




\begin{abstract}
Perturbative methods are attractive to describe the electronic structure of molecular systems 
because of their low-computational cost and systematically improvable character. 
In this work, a two-step perturbative approach is introduced combining multi-state Rayleigh-Schrödinger (effective Hamiltonian theory) and state-specific Brillouin-Wigner schemes to treat
degenerate configurations and yield an efficient evaluation of multiple energies.
The first step produces model functions and an updated definition of the perturbative partitioning of the Hamiltonian.
The second step inherits the improved starting point provided in the first step, enabling then faster processing of the perturbative corrections for each individual state.
The here-proposed two-step method is exemplified on a model-Hamiltonian of increasing complexity.

\end{abstract}

 \maketitle
 \ioptwocol

%
%
%

%

%

\section{Introduction}

The accurate description of electronic correlation in molecular systems remains a long-standing challenge in quantum chemistry.
To describe such non-trivial effects, a Full Configuration-Interaction (FCI) decomposition of the wavefunction provides the best qualitative result  within a finite size basis set~\cite{helgaker2014molecular}.
However in practice, FCI is known to suffer from the so-called \textit{exponential wall} problem which restricts its application to very small-sized systems.
To circumvent this problem, the Complete-Active-Space (CAS) approximation is usually convoked in order to reduce
the wavefunction expansion.
The calculation becomes then numerically tractable (within a certain size limit~\cite{levine2020casscf,nakatani2017density,li2016combining}), but it comes with the 
limitations resulting from a reduced  space. 
To recover part of the missing correlation contributions, different strategies have been developed. 
One of them consists in using orbital optimization techniques (\textit{i.e.} optimal rotations of the single-particle-functions basis).
This idea forms the basis of the famous CAS Self-Consistent-Field (CASSCF) method~\cite{roos1980complete,siegbahn1980comparison,siegbahn1981complete} which is considered as a reference tool to describe strongly correlated chemical systems.
In this variational approach, orbital rotations and CAS expansion are employed in order to re-encode a large part of the electronic correlation within the active space.
This contribution is commonly called \textit{"static correlation"}.
Orbital optimization techniques have been known for a while, but they have recently received renewed attention from the community for the development of new methodologies~\cite{ding2023quantum,yalouz2021state,mizukami2020orbital,mahler2021orbital,ammar2023bi,yalouz2023orthogonally,yao2021orbital,barca2018simple, wouters2014density,keller2015efficient,liu2013multireference}.
In practice, because of the active space approximation, it is known that CASSCF-like calculations fail to fully recover
electronic correlation.
This energetic contribution
is named \textit{dynamical correlation} as it stems from 
many-electron excitations changing the mean-field
picture provided by the CAS picture. 
For strongly correlated systems,  dynamical correlation is decisive to reach spectroscopic accuracy in electronic structure calculations~\cite{chang2012multi,amor1998size,roseiro2023modifications,vela2017electron}.
As a solution to this problem, perturbation theory (PT)~\cite{lindgren2012atomic} provides a solid framework that can in principle be systematically improved. 
In a PT scheme, the full Hamiltonian 
$\hat{H}$
is split into two parts, namely a zeroth order  Hamiltonian $\hat{H_0}$ 
and a perturbation $\hat{V}$, such that $\hat{H} = \hat{H_0} + \hat{V}$. 
In practice, not only must $\hat{H_0}$ include the  
dominant contributions, but it must be simple enough to produce 
reference eigenstates $\lbrace \ket{\alpha} \rbrace $ and eigenvalues $ \lbrace E_\alpha \rbrace $.
Depending on the system's complexity (e.g. presence of open-shells) and 
the inspected quantities, such reference Hamiltonian
might be obtained either from Hartree-Fock (single-reference) or 
CASSCF  (multi-reference) methods.
Whatever the subsequent perturbation scheme, the presence of quasi-degenerate states 
calls for particular attention with the appearance of singularities
and possible intruder states (as in CASPT2~\cite{roos1982simple,andersson1990second}). 
A clear-cut separation between the model space (e.g. the CAS) and the orthogonal space (formed by electronic configurations outside the CAS) might be blurred as soon as the perturbation is turned on.
In this case, vanishing energy denominators lead to divergence
in the expansion series and the application of
quasi-degenerate perturbation theories may become questionable.
Other schemes such as NEVPT2~\cite{angeli2001introduction,angeli2001n,angeli2002n} have also been introduced with the appealing features that it can avoid such a divergence problem.

Many PT developments are inspired from the Brillouin-Wigner (BW) or Rayleigh-Schrödinger (RS) schemes (see for example Refs.~\cite{li2023toward,yi2019multireference,hubavc2010brillouin, hubavc1994size, wilson2003brillouin}, and references within).
The key advantage of the BW approach lies in
the absence of the intruder states problem whereas
multi-reference methods, such as RS,  suffer
from their possible appearance~\cite{lindgren2012atomic}. 
Still, the BW perturbation theory is less widely used than the RS scheme, possibly due to the
fact that the energy correction directly 
depends on the exact, and unknown, energy
and its inadequate scaling
with the number of particles.
However, its straightforward extension to 
any high orders and its ability to produce state-specific
energy corrections make it very appealing.

Based on both BW and RS perturbation theories, we here propose 
a perturbative scheme combining the strengths of the two approaches to
reach state-specific energy corrections in two steps. 
In the first step, the RS method is used to generate a perturbative effective Hamiltonian that
can provide model functions for a given model space.
In a second step, such an effective Hamiltonian is
taken as an updated zeroth order Hamiltonian
along with a modified perturbation to define a state-specific BW expansion.
Following this two-step approach, one accounts for quasi-degeneracies (within RS) and then concentrates on a set of problems where a single energy level is treated at a time (within BW).
The Rayleigh-Schrödinger-Brillouin-Wigner (RSBW) method is first described and then applied to parameterized model-systems of increasing complexity.
The method is compared to other schemes to stress the convergence improvements and possible limitations.

 \begin{figure*} 
     \centering
     \includegraphics[width=14cm]{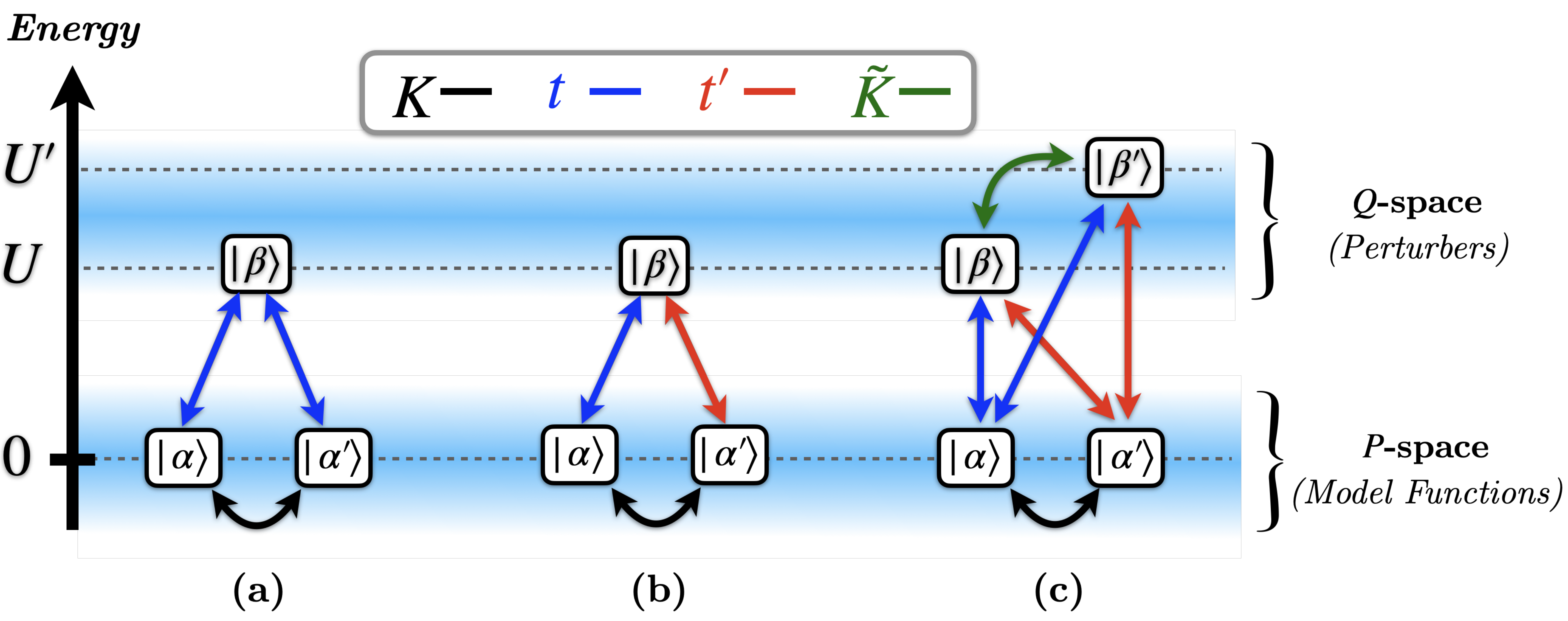}
     \caption{Schematic representation of the three different cases  of interactions considered between the model and the perturber spaces. (a) Symmetric interaction between two degenerate model functions  and a single perturber. (b) Asymmetric interactions between two degenerate model functions  and a single perturber. (c) Asymmetric interactions between two degenerate model functions  and two non-degenerate perturbers.
     } 
     \label{fig:scheme_models}
 \end{figure*}

\section{Two-step perturbative method: Rayleigh-Schrödinger and Brillouin-Wigner }
The leading Brillouin-Wigner (BW) and Rayleigh-Schrödinger (RS) perturbation approaches~\cite{lindgren2012atomic} 
rely on the definition of an appropriate  model space $P$
constructed on the model functions
$\Bqty{\ket{\alpha}}$, eigenfunctions of the unperturbed Hamiltonian $\hat{H_0}$:
\begin{equation}
    \hat{H}_0 | \alpha \rangle = E_{\alpha} | \alpha \rangle 
\end{equation}
The orthogonal  $Q$-space is spanned by the so-called $\{\ket{\beta}\}$
perturbers with energies $E_{\beta} = \mel{\beta}{\hat{H}_0}{\beta}$.
In a quasi-degenerate picture, the impact of the perturbation 
must be evaluated within the $P$-space first. 
In practice, the construction of an effective Hamiltonian $\hat{H}_{\mbox{\tiny eff}}$ that operates in
the model space
allows one to produce the associated effective 
model functions $\{\ket{\Psi^{\mbox{\tiny RS}}_i}\}$ (eigenvectors of $\hat{H}_{\mbox{\tiny eff}}$ in the $P$-space).
At second order, 
the effective Hamiltonian matrix elements can be derived from the generalized Bloch equation:
\begin{equation}
    \begin{split}
    \mel{\alpha}{\hat{H}_{\mbox{\tiny eff}}^{(2)}}{\alpha'} &= \delta_{\alpha \alpha'} E_{\alpha'} \\
    &+ \mel{\alpha}{\hat{V}}{\alpha'} \\
    &+ \sum_{\beta}^\text{$Q$-space} \frac{\mel{\alpha}{\hat{V}}{\beta}\!\!\mel{\beta}{\hat{V}}{\alpha'}}{E_{\alpha'} - E_{\beta}} 
    \end{split}
    \label{eq:Heff}
\end{equation}
It is well-known that the effective Hamiltonian
is not hermitian.
Such deviation arises from the summation in Eq.~(\ref{eq:Heff}).
Since the latter is expected to contribute less (\textit{i.e.} second order) 
than the first and second terms,
one can easily remedy this issue
by averaging over the off-diagonal elements .
Despite its approximate character, the diagonalization
of the second order effective Hamiltonian 
$\hat{H}_{\mbox{\tiny eff}}^{(2)}$ generates qualitative
effective model functions $\{\ket{\Psi^{\mbox{\tiny RS}}_i}\}$ and associated eigenvalues $\{E^{\mbox{\tiny RS}}_i\}$
resulting from the
mixing of the  $\{\ket{\alpha}\}$
through the perturbation $\hat{V}$.
Importantly, this rotation within the $P$-space leaves
the  orthogonal states and the \textcolor{black}{couplings within the $Q$ space}  unchanged.
From this first step, a different PT partitioning of the 
Hamiltonian is introduced such that $\hat{H} = \hat{H}^{\mbox{\tiny RS}} + \hat{W}$, with 
\begin{equation}
    \hat{H}^{\mbox{\tiny RS}} = \sum_{i}^\text{$P$-space} E^{\mbox{\tiny RS}}_i \ketbra{\Psi^{\mbox{\tiny RS}}_i}{\Psi^{\mbox{\tiny RS}}_i} +   \sum_{\beta}^\text{$Q$-space} E_{\beta} \ketbra{\beta}{\beta}
    \label{eq:HRS_def}
\end{equation}
$\hat{W}$ is an improved perturbation operator in the sense that it should carry fewer significant contributions as compared to $\hat{V}$.
Regarding the perturbation treatment, the expected improvement can be quantified by evaluating the ratio:
\begin{equation}
\begin{split}
\tau = \frac{\rho_{\mbox{\tiny RSBW}}}{\rho_{\mbox{\tiny RS}}}, 
\end{split}
\label{eq:tau_definition}
\end{equation}
where $ \rho_{\mbox{\tiny RS}}  $ and  $ {\rho_{\mbox{\tiny RSBW}}}$ quantify the perturbation 
contributions (respectively $\hat{V}$ and $\hat{W}$) over the reference unperturbed Hamiltonians (respectively $\hat{H}_0$ and $\hat{H}^{\mbox{\tiny RS}}$) at each step of the RSBW method, such that
\begin{equation}
\begin{split} 
\rho_{\mbox{\tiny RS}} = \frac{\vert \vert \hat{V} \vert \vert}{\vert \vert \hat{H}_0 \vert \vert},
\hspace{0.1em} \mbox{  and  } \hspace{0.1em}
\rho_{\mbox{\tiny RSBW}} = \frac{\vert \vert \hat{W} \vert \vert}{\vert \vert \hat{H}_{\mbox{\tiny RS}} \vert \vert},
\end{split}
\label{eq:rho_definition}
\end{equation}
with the notation 
\begin{equation}
\begin{split}
\vert \vert \hat{O} \vert \vert = 
\left(\Tr\, [\hat{O}^\dagger \hat{O}]  \right)^{1/2}
\end{split}
\label{eq:norm}
\end{equation}
for the norm of a generic operator $\hat{O}$ evaluated over the full $P+Q$ space~\cite{golub2013matrix}.
The absence of degenerate states makes the BW theory particularly appealing with
a straightforward expansion of the exact state energy $E_i$:
\begin{equation}
    \begin{split}
    E_i =  \mel{\Psi_i}{\hat{H}}{\Psi_i} &+ 
    \sum_{n >1} \mel{\Psi_i} {\hat{W} \hat{\Omega}^{\left(n -1 \right)}} {\Psi_i}
    \end{split}
    \label{eq:BW_general1}
\end{equation}
with
\begin{equation} 
    \hat{\Omega}^{\left(n \right)} = 
   \left( \sum_{j \neq i} \frac{\ketbra{\Psi_j}}{E_i - \hat{H}^{\mbox{\tiny RS}}} \hat{W} \right)^{\!\!n}. 
    \label{eq:BW_general2}
\end{equation}
In practice, the summation in  Eq.~(\ref{eq:BW_general1}) is truncated and the exact energy $E_i$
is approximated as 
$E^{\mbox{\tiny RSBW}}_i$ by setting the energy denominators 
to $E^{\mbox{\tiny RSBW}}_i - E^{\mbox{\tiny RS}}_j$
and $E^{\mbox{\tiny RSBW}}_i - E_{\beta}$:
\begin{equation}
    \begin{split}
    E^{\mbox{\tiny RSBW}}_i &=  \mel{\Psi_i}{\hat{H}}{\Psi_i} \\ &+ 
    \sum_{j \neq i}^\text{$P$-space} \frac{ \mel{\Psi_i}{\hat{W}}{\Psi_j} \!\! \mel{\Psi_j}{\hat{W}}{\Psi_i} }
    {E^{\mbox{\tiny RSBW}}_i - E^{\mbox{\tiny RS}}_j} \\ 
    &+ \sum_{\beta}^\text{$Q$-space} \frac{\mel{\Psi_i}{\hat{W}}{\beta} \!\! \mel{\beta}{\hat{W}}{\Psi_i} }
    {E^{\mbox{\tiny RSBW}}_i - E_{\beta}} \\
    &+ \ldots
    \end{split}
    \label{eq:BW}
\end{equation}

Importantly, the first summation in  Eq.~(\ref{eq:BW})   accounts for the mixing of the
model functions through the modified perturbation $\hat{W} = \hat{H} -\hat{H}^{\mbox{\tiny RS}}$.
Not only does the RS treatment lift the quasi-degeneracy (see the energy denominators in Eq.~(\ref{eq:BW}), $E^{\mbox{\tiny RSBW}}_i - E^{\mbox{\tiny RS}}_j \approx E^{\mbox{\tiny RS}}_i - E^{\mbox{\tiny RS}}_j$)
but it also involves  $\hat{W}$ (instead of  $\hat{V}$). 
The $E^{\mbox{\tiny RSBW}}_i$ values can
be obtained from iterations on Eq.~(\ref{eq:BW}), 
including all terms at any arbitrary order.
In the following, we will refer to the iter-RSBW method.
When limited to first and second order contributions, the 
resulting quadratic equation can be further simplified
by assuming $E^{\mbox{\tiny RSBW}}_i \approx  E^{\mbox{\tiny RS}}_i$
in the energy denominators.
This less demanding approach  
which should qualitatively validate the partitioning
is named RSBW method.

By concentrating part of the perturbation in the redefinition 
of the zeroth-order Hamiltonian $\hat{H}_{\mbox{\tiny RS}}$ (given in Eq.~(\ref{eq:HRS_def})), 
one can expect
(\textit{i}) an improved convergence of the BW series
as measured by $\tau <1$,
(\textit{ii}) a reduction of the number of iteration steps,
and (\textit{iii}) a reduced size-consistency error.
In the following sections, the RSBW and iter-RSBW methods that combine RS and BW 
perturbation theories are
exemplified on a model Hamiltonian built on a strictly degenerate
two-dimensional $P$-space. 
The dimension of the orthogonal $Q$-space is varied to 
account for configuration interaction.
Along our inspections, the quantities defined in Eq.~(\ref{eq:rho_definition})
are  evaluated to stress the method's relevance and applicability.

\section{Numerical results on model Hamiltonians}

The energy splittings and the structure of the
wavefunctions were examined on systems ruled by model Hamiltonians. To be demonstrative and to limit the number of parameters,  we concentrated on
a two-dimensional  strictly degenerate model space spanned by the reference 
functions $\ket{\alpha}$ and $\ket{\alpha'}$. The unperturbed Hamiltonian
energies  
$\mel{\alpha}{\hat{H}_0}{\alpha} = \mel{\alpha'}{\hat{H}_0}{\alpha'}$
are set to zero whilst the perturbation within the $P$-space is
$\mel{\alpha}{\hat{V}}{\alpha'}$. 
The model space interacts with
the orthogonal space  
through $V_{\alpha \beta} = \mel{\alpha}{\hat{V}}{\beta}$
and the perturbers $\left\{\ket{\beta}\right\}$ lie $E_{\beta}=\mel{\beta}{\hat{H}_0}{\beta} >0 $  
higher in energy.
Despite its simplicity, this picture is representative of the emblematic "two electrons in two orbitals" problem (\textit{i.e.} CAS[2,2]) encountered in bond dissociation processes and exchange coupling
constants calculations. The $Q$-space affords for fluctuations with
respect to a mean field description (e.g. inclusion of charge transfer configurations) and non-zero
$V_{\beta \beta'}=\mel{\beta}{\hat{V}}{\beta'}$ terms account for 
configuration interaction processes. Too small  an $E_{\beta}$ value relatively to the energy splitting 
within the model space ($\propto \vert \mel{\alpha}{\hat{H}}{\alpha'} \vert$) makes $\ket{\beta}$ an intruder state
that should be included in the model space. Evidently, the partitioning 
should then be reconsidered at the relatively small cost of an enlarged $P$-space. 
Without loss of generality, such a scenario will be avoided in the following by restricting the ranges of parameter variations.

For preliminary inspections, 
we considered a symmetric system characterized by a strictly degenerate model space
coupled to a one-dimensional $Q$-space (see Figure~\ref{fig:scheme_models}(a)).
This is the traditional one-band model used to interpret the physical mechanisms
that govern the magnetic coupling.
For this reason, the standard notations are introduced  for the matrix elements
with $t = V_{\alpha \beta} = V_{\alpha' \beta}$,  $K = \mel{\alpha}{\hat{V}}{\alpha'}$ 
and $U =\mel{\beta}{\hat{H}_0}{\beta} >0$.
From symmetry considerations, the exact
model functions are the in-phase  and out-of-phase 
linear combinations of
the reference functions, $\ket{\Psi_g} =  \frac{1}{\sqrt{2}} \left(\ket{\alpha} + \ket{\alpha'} \right)$
and $\ket{\Psi_u} =  \frac{1}{\sqrt{2}} \left(\ket{\alpha} - \ket{\alpha'} \right)$.
For the perturbative treatments to be 
valid, the absolute couplings  $\vert t \vert$ between the $P$- and $Q$-spaces must be smaller than the energy separations between their respective states. Qualitatively, the splitting within the model space being \textit{ca.} 2$\vert K \vert$, such conditions require $U \pm K > \vert t \vert$, 
setting the ranges of parameters we will use for numerical inspections.
%
Following the Bloch theory~\cite{lindgren2012atomic}, the second order effective Hamiltonian can be written as a dressed matrix :
\begin{equation}
\begin{split}
\hat{H}_{\mbox{\tiny eff}}^{(2)}  = 
\begin{pmatrix}
    -\frac{t^2}{U} & K  - \frac{t^2}{U}  \\
    K  - \frac{t^2}{U}  & -\frac{t^2}{U} \\
\end{pmatrix}
\end{split}
\label{eq:RS_matrix}
\end{equation}
The  eigenvalues
$E^{\mbox{\tiny RS}}_g = K - 2{t^2}/{U}$ and $E^{\mbox{\tiny RS}}_u = - K$ 
can be compared to the undressed model space energies 
$\pm K$.
In general, $E^{\mbox{\tiny RS}}_g$ and $E^{\mbox{\tiny RS}}_u$  
are \textit{a priori} approximations of the exact energies of the full Hamiltonian $\hat{H}$.
However, in the case under study, 
$E^{\mbox{\tiny RS}}_u$ is the exact energy
for symmetry reasons ($\mel{\Psi_u}{\hat{W}}{\beta} = 0$).
In contrast, the  $\ket{g}$ state energy
is modified through the 
second-order BW expansion (see Eq.~(\ref{eq:BW})). 
Remembering that $\mel{\Psi_g}{\hat{W}}{\Psi_u} = 0$, no contribution arises from the
mixing within the model space. 
Besides, the $Q$-space being one-dimensional and 
$\mel{\beta}{\hat{W}}{\beta}  = 0$, 
the BW expansion ends at second-order with identically null higher-order terms :
\begin{equation}
\begin{split}
E^{\mbox{\tiny RSBW}}_g =  K \ + \
& \frac{2t^2}{E^{\mbox{\tiny RSBW}}_g - U}
\end{split}
\label{eq:BW_degenere}
\end{equation}
This quadratic equation is the one we would obtain from the exact diagonalization of the full Hamiltonian
in the $\left\{\ket{\Psi_g}, \ket{\beta} \right\}$ sub-space. 
This can be understood as follows: first, the symmetry of the problem fully defines the  model functions $\ket{\Psi_g}$ and  $\ket{\Psi_u}$,
which makes any further expansion of the 
effective Hamiltonian unnecessary.
Then, the BW treatment being exact at second order, Eq.~(\ref{eq:BW_degenere}) produces the exact energy. 
To avoid any iteration or quadratic equation resolution
, $E^{\mbox{\tiny RSBW}}_g$ can be approximately 
evaluated by setting  $E^{\mbox{\tiny RSBW}}_g $ to $E^{\mbox{\tiny RS}}_g = K - 2 {t^2}/{U} $ in the energy denominator (\textit{i.e.} RSBW method).
 \begin{table}[!ht]  
  \centering
  \begin{tabular}{c|c|c}
    \toprule
     RS       & BW       & RSBW   \\ \midrule
     $28\%$   & $6.7\%$  & $3.9\%$    \\ \bottomrule
  \end{tabular}
  \caption{{Relative errors of the ground state energy of a symmetric system ($t = t'$, Figure~\ref{fig:scheme_models}(a)) using the RS, BW and RSBW. $K = -1$, $t' = -1$ and $U=2$ in $\lvert t \rvert$  unit.}}
  \label{table:1}
\end{table} 

Table~\ref{table:1} compares the relative errors for the ground state energy evaluated using the RS and RSBW approaches, 
and a BW-like correction to the 
 $\ket{\Psi_g}$ state energy,
$E^{\mbox{\tiny BW}}_g =  K  + 
{2t^2}/({K - U})$.
Despite its simplicity, the model 
supports the two-step RSBW method
which not only recovers the exact solution (see Eq.~(\ref{eq:BW_degenere})), but
also suggests a costless energy evaluation 
(3.9\% relative error).
The improved perturbation is measured by $\tau = 0.6$
(see Eq.~(\ref{eq:rho_definition}) with $K=-1$, $t' = -1$ and $U = 2$ in $\lvert t \rvert$ unit), highlighting the benefits brought by the $\hat{H} = \hat{H}^{\mbox{\tiny RS}} + \hat{W}$ splitting.

\begin{figure}[t]
     \centering
     \includegraphics[width=8cm]{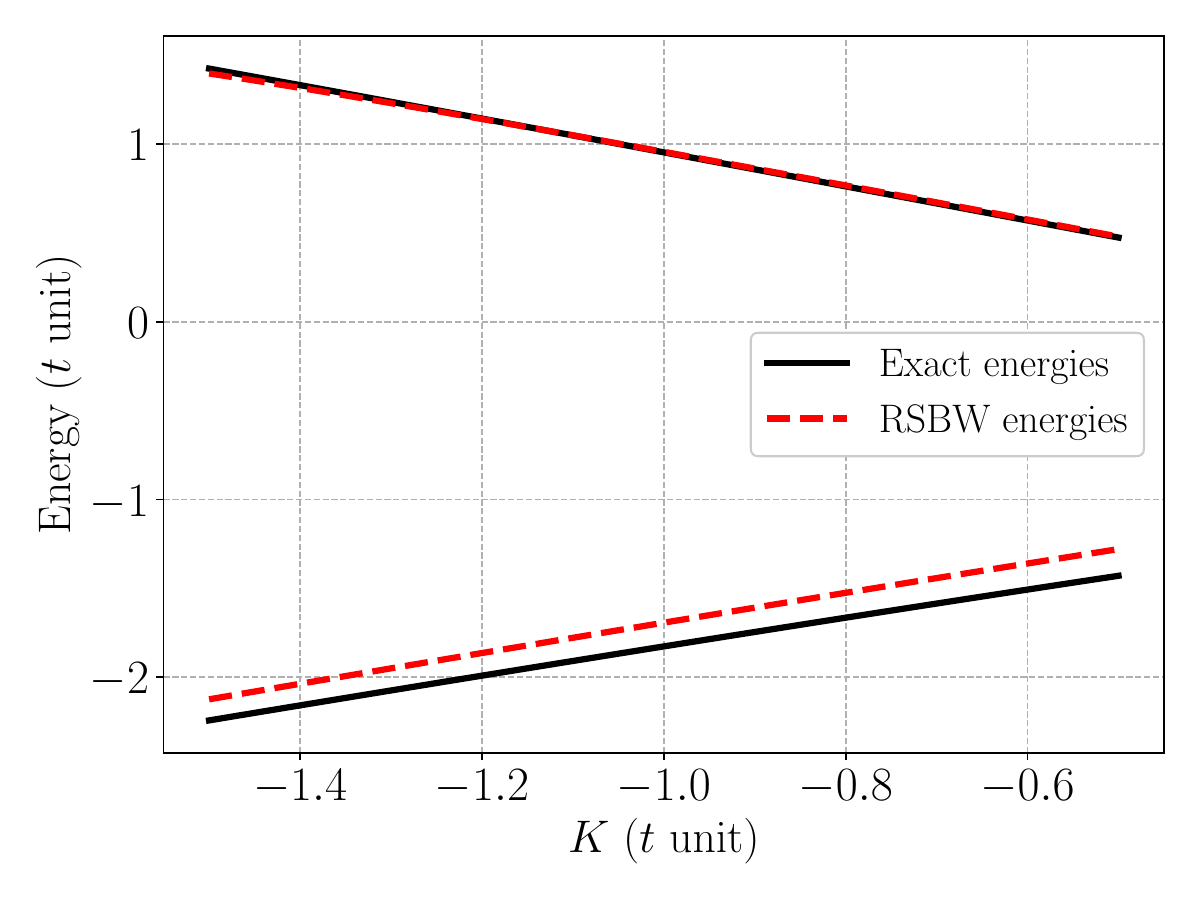}
     \caption{{Ground and excited states energies $E^{\mbox{\tiny RSBW}}_1$ and $E^{\mbox{\tiny RSBW}}_2$ of an asymmetric system
     (Figure~\ref{fig:scheme_models}(b)) obtained from the second-order RSBW approach as a function of $K$ (all energies $E^{\mbox{\tiny RSBW}}_i$ in the energy denominators of Eq.~(\ref{eq:BW}) are set to $E^{\mbox{\tiny RS}}_i$). The exact energies are given for comparison. A single perturber lies $U=2$ higher above the degenerate model space, $t' = -1.5$, and all energies are in $\lvert t \rvert$ unit.}}
     \label{fig:RSBW}
 \end{figure}

To complement these preliminary inspections, the method was then applied to a degenerate model-space
interacting with a single perturber with $t = V_{\alpha \beta} \neq  V_{\alpha' \beta} = t'$ (see Figure~\ref{fig:scheme_models}(b)).
The dressed matrix given in Eq.~(\ref{eq:RS_matrix}) was modified accordingly, and subsequently diagonalized
to produce the model functions $\ket{\psi_1} = \mbox{cos} ( \theta)  \ket{\alpha} + \mbox{sin} ( \theta)  \ket{\alpha'}$
and $\ket{\psi_2} = -\mbox{sin}  (\theta)  \ket{\alpha} + \mbox{cos}  (\theta)  \ket{\alpha'}$
and energies 
$E^{\mbox{\tiny RS}}_1$ and $E^{\mbox{\tiny RS}}_2$, 
with 
\begin{equation}
    \mbox{tan} (\theta) = \displaystyle \frac{U \sqrt{\Delta} + t^2 - t'^2}{U K - t t'}
\end{equation} 
and 
\begin{equation}
    \Delta = \pqty{\frac{t^2 + t'^2}{2U}}^2 + K \pqty{K - 2 \frac{t t'}{U}}.
\end{equation}
The absence of symmetry (\textit{i.e.} $\theta \neq \frac{\pi}{4}$) gives rise to a non-zero 
$\mel{\psi_1}{\hat{W}}{\psi_2}$ matrix element.
Even though the BW expansion contains 
higher-order terms, it was
truncated  at second order for comparison
purposes, and the exact energies $E_i$ 
were approximated as $E^{\mbox{\tiny RSBW}}_i$.
Following the RSBW method (\textit{i.e.} no iteration),  the resulting quadratic
equations were further simplified  by setting the
energy denominators to $E^{\mbox{\tiny RS}}_i - E^{\mbox{\tiny RS}}_j$ and $E^{\mbox{\tiny RS}}_i - U$
($i,j=1,2$) in Eq.~(\ref{eq:BW}).
From Figure~\ref{fig:RSBW}, 
reasonably good agreement between
the approximated $E^{\mbox{\tiny RSBW}}_1$
and $E^{\mbox{\tiny RSBW}}_2$ values and
the exact ones
is reached 
at the negligible cost of the redefinition of the zeroth-order Hamiltonian.
Despite the similarity between the coupling value
and the energy separation with the perturber
($t / \left(U + K\right) \sim  -1$),
the excited state energy is reproduced with a relative error smaller than 2\% whatever the $K$ value.
However, a stronger deviation  is observed for the  ground  state energy, a failure inherent to the BW approach that calls for  a self-consistency procedure.
 \begin{figure}[t]
     \centering
     \includegraphics[width=8cm]{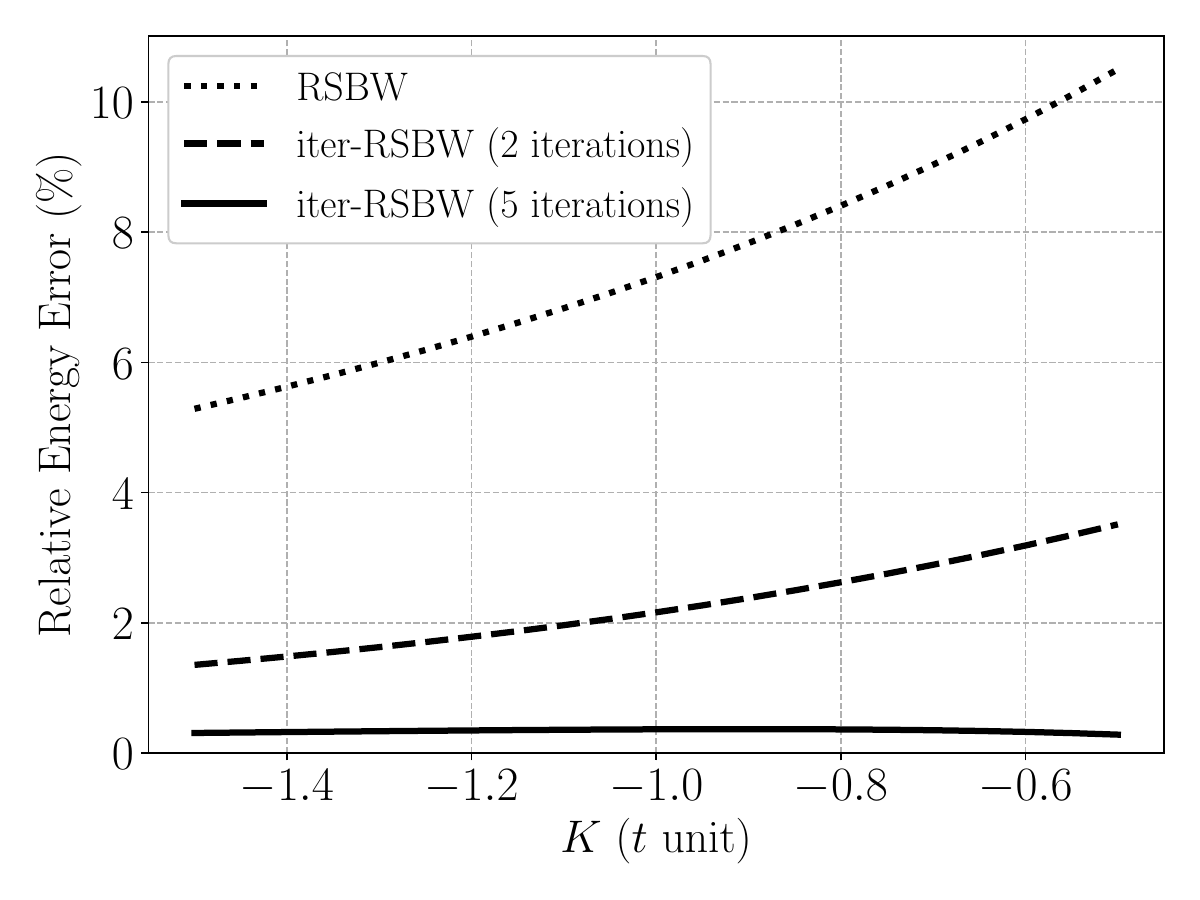}
     \caption{
     {Relative error for the ground state energy of an asymmetric system
     (Figure~\ref{fig:scheme_models}(b)) obtained from the RSBW and iter-RSBW approaches as a function of $K$. A single perturber lies $U=2$ higher than the model space, $t' = -1.5$, all energies are in $\lvert t \rvert$ unit.}} 
     \label{fig:RSBW_iterated}
 \end{figure}
Along the iter-RSBW method, the relative error becomes negligible (smaller than 0.4\%) after a limited number of iterations (see Figure~\ref{fig:RSBW_iterated}).
The  RSBW   procedure is improved
as supported by the $\tau$ values reported in Table~\ref{table:2}. 
 \begin{table}[!ht]  
  \centering
  \begin{tabular}{c|c|c|c}
    \toprule
    $K$       & -2       & -1.5 & -1    \\ \midrule
    $\tau$    & 0.3      &  0.5 &  0.6   \\ \bottomrule
  \end{tabular}
  \caption{{$\tau$ values as defined in Eq.~(\ref{eq:tau_definition}) stressing the improved convergence. $t' = -1.5$ and $U = 2$  in $\lvert t \rvert$  unit.}}
  \label{table:2}
\end{table} 
Let us stress that whatever the number of iterations, the strict BW approach
limited to second-order fails to recover the 
transition energy $\Delta E = E_2-E_1$
with relative errors larger than 7\%.
Even though analytical solutions can be extracted, our intention is to stress the
rapid convergence upon iterations,
with a natural extension to higher orders
in the BW  expansion.
For more realistic problems in quantum chemistry, the dimension
of the $Q$-space would be larger. However, the size of the second order effective Hamiltonian (see Eq.~(\ref{eq:RS_matrix}))
remains unchanged and the decisive  BW treatment is easily implemented with an improved convergence.

\begin{figure}[t]
     \centering
     \includegraphics[width=8cm]{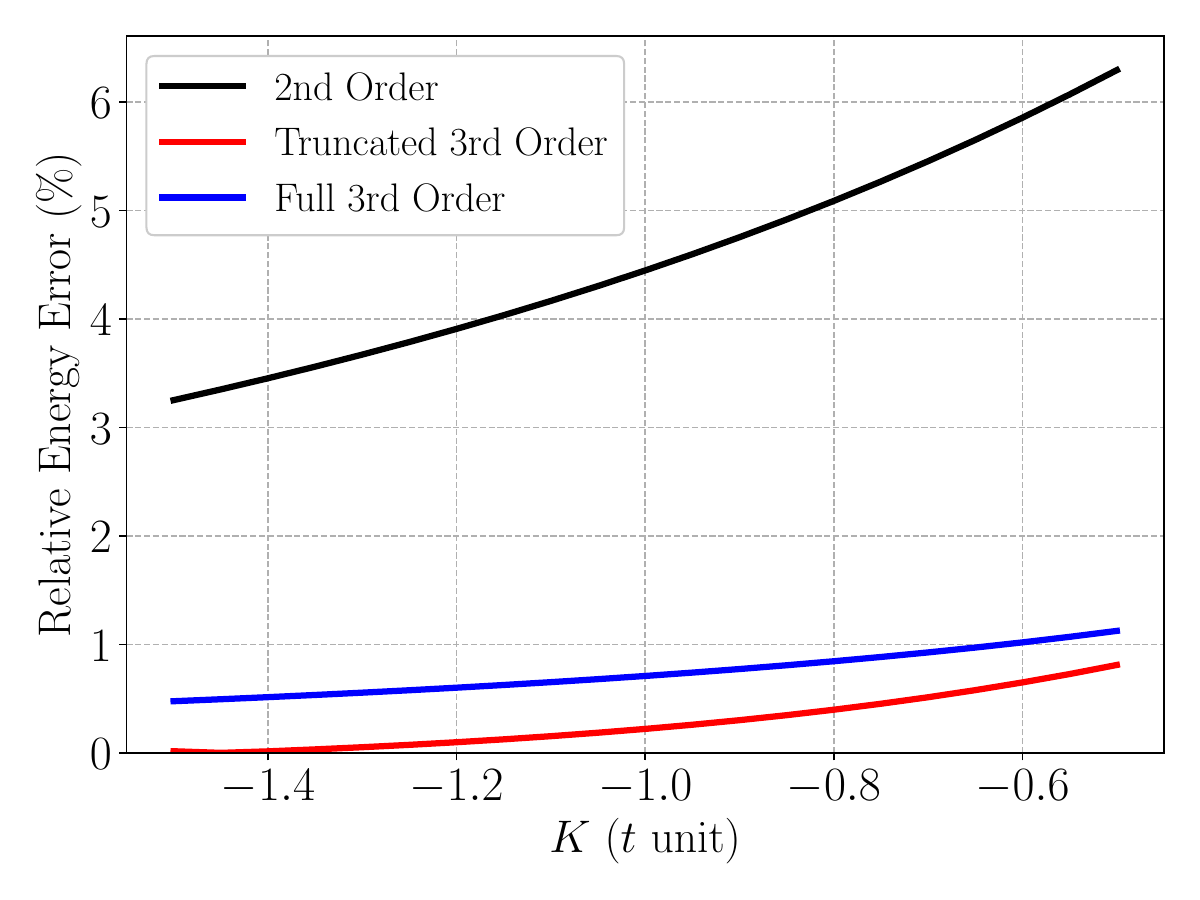}
     \caption{{Relative error for the ground state energy of an asymmetric system (Figure~\ref{fig:scheme_models}(c)) obtained from the second order, truncated, and third order iter-RSBW approaches as a function of $K$. Two interacting perturbers, $\Tilde{K} = -0.5$, lie $U=2$ and $U'=3$ higher above the model space, $t' = -1.5$.  All energies are in $\lvert t \rvert$ unit.}} 
     \label{fig:RSBW_CI_iterated}
 \end{figure}

Finally,  non-zero matrix elements
$\Tilde{K} = \mel{\beta}{\hat{W}}{\beta'}$
may couple perturbers within the $Q$-space
in configuration interaction calculations
(see Figure~\ref{fig:scheme_models}(c)). As a consequence, additional higher-order contributions are likely to emerge
and can be easily evaluated due to the simplicity of the BW expansion.
The same procedure was followed by first generating the model functions
issued from the diagonalization of the second-order effective Hamiltonian.
As expected, the second order iter-RSBW method fails to reproduce the exact energies (see Figure~\ref{fig:RSBW_CI_iterated}).
Thus, the BW series was expanded up to 
third order contributions and iterations were carried out to evaluate the iter-RSBW states energies.
Particular attention was paid to the 
third-order contributions involving
$\Tilde{K}$. The latter can be easily isolated in the 
BW expansion, excluding all the other third-order terms.
After iterations, the truncated third order evaluation
of the energies are referred to as truncated iter-RSBW.
Interestingly,
most of the missing dynamical correlation is faithfully recovered along this procedure.
Such observation suggests that the iter-RSBW method allows one to concentrate 
the numerical effort on specific contributions.
Let us finally mention that the transition energy $\Delta E$
evaluated from the truncated iter-RSBW is in excellent agreement
(error smaller than 0.5\%) with the exact value.
The redefinition of the zeroth order Hamiltonian improves the BW convergence
($\tau = 0.8$ for $K = -1$ in $\abs{t}$ unit)
and sets a hierarchy in the third (and beyond) order contributions.

\section{Conclusion}
The method we proposed  combines the 
Rayleigh-Schrödinger (RS) and Brillouin-Wigner (BW) perturbation theories.
The objective is to take advantage of both methods and to progressively include the perturbation effects.
The relevance of this two-step
strategy is evaluated on a series of model Hamiltonians.
In the simplest symmetrical case, the analytical solution is
retrieved.
The construction of the second order effective Hamiltonian (model space) in the RS framework 
produces model functions. Even though this first step cannot safely account for 
the transition energy, a modified splitting of the
Hamiltonian emerges with a redefinition of the
perturbation.
In the second step, a systematic and costless order-by-order BW treatment 
can be performed on each individual state (state-specific) and leads to accurate energy
estimations.   
It is shown that the convergence of the BW energy expansion is improved, as featured by
a reduction of the perturbation
norm with respect to the zeroth order Hamiltonian. 
As soon as the perturbers interact in the orthogonal space (traditional CI structure),
it is suggested that (\textit{i}) a second order BW treatment fails to recover
dynamical correlation effects,
 and (\textit{ii}) the leading third order contributions involve
couplings within the orthogonal space.
An alternative approach would follow an iterated RS approach
where the second step would consist in a redefinition of the
model Hamiltonian.
Depending on the size of the model space, the construction of successive effective Hamiltonians might 
become numerically demanding whereas the BW expansion is easily implemented at any order.
Such extension and practical implementation for quantum chemistry Hamiltonians are left for future developments.

\section{Acknowledgments}
This work was supported by the Interdisciplinary Thematic Institute SysChem via the IdEx Unistra (ANR-10-IDEX-0002) within the program Investissement d’Avenir.
The authors wish to thank Prof. C. Angeli for helpful discussions, and L. Petit for preliminary inspections and for furnishing the basis of the program used to generate the results presented in this article.
 
\section*{References}
\bibliographystyle{iopart-num} 
\bibliography{biblio}

\end{document}